\documentclass[aps,prd,preprint,groupedaddress]{revtex4}
\usepackage{amssymb}
\usepackage{graphicx}
\usepackage{bm}
\usepackage{amsmath}

\begin{document}


\title{Vacuum polarization of massive spinor fields \\
in static black-string backgrounds}
\author{Owen Pavel Fern\'{a}ndez Piedra}
\email{opavel@ucf.edu.cu }
\affiliation{Departamento de F\'{i}sica y Qu\'{i}mica, Universidad de Cienfuegos, Carretera a Rodas, Cuatro Caminos, s/n. Cienfuegos, Cuba.}

\author{Alejandro Cabo Montes de Oca}
\email{cabo@icmf.inf.cu}
\affiliation{Grupo de F\'{i}sica Te\'{o}rica, ICIMAF, Calle E  No. 309, esq. a 15 Vedado, C. Habana,Cuba.}

\date{\today}

\begin{abstract}

 \noindent The renormalized mean value of the quantum Lagrangian and
 the corresponding components of the
Energy-Momentum tensor for  massive spinor fields coupled  to an arbitrary gravitational field configuration having cylindrical symmetry  are analytically evaluated using the Schwinger-DeWitt
approximation, up to second order in the inverse mass value. The general results are employed to explicitly derive compact analytical expressions for the quantum mean Lagrangian and Energy-Momentum
tensor in the particular background of the Black-String space-time.
\end{abstract}

\pacs{04.62.+v,04.70.Dy}

\maketitle

\section{Introduction}

In the absence of a full theory of quantum gravity, semiclasical gravity or Quantum Field theory in curved space-time is a well established physical theory that help us to know what are the expected
behavior of gravitational system under the influence of the interaction between it and matter fields that obeys the laws of quantum theory. The research on this subject have received a great impulse
since the Hawking discovery of black holes radiation \cite{Hawking}.

One of the most important physical quantities to be determined in semiclassical gravity is the vacuum expectation value of the stress energy tensor \(\langle T_{\mu}^{\nu}\rangle_{ren}\) of the
quantum field . This quantity enters as a source in the semiclassical Einstein equations, which determines the changes in the gravitational background field due to its interaction with the quantum
one (the backreaction). Up to know, there exist many works related to the calculation of the components of the renormalized stress tensor by different approaches. The main difficulty in the problem
of calculate the components of \(\langle T_{\mu}^{\nu}\rangle_{ren}\)  is the dependence of this quantity on the metric tensor of the background gravitational field. For this reason it is impossible
to have an exact analytical formula for this object. Except for some exact expressions for this quantity obtained for very special spacetimes , on which quantum matter fields propagates, and for
boundary conditions with a high degree of symmetry \cite{dowker,browncassidy,bunchdavies,allenfolacci,kirsten}, the majority of the techniques developed rest on approximate methods to build the
energy momentum tensor or numerical computations of \(\langle T_{\mu}^{\nu}\rangle_{ren}\) and the mean-squared field \(\langle\varphi^{2}\rangle_{ren}\)
\cite{howardcandelas,candelas,fawcet,jensen12,ADL,anderson,demello}. One of the developed techniques is the Schwinger-De Witt expansion, that can be used to investigate effects like the vacuum
polarization of massive fields in curved backgrounds. The Schwinger-DeWitt approximation results from an expansion of the effective action in powers of the inverse mass of the quantum field, and is
valid to use it whenever the Compton's wavelenght of the field is less than the characteristic radius of curvature \cite{frolov, DeWitt, avramidi,barvinsky, matyjasek, matyjasek1}.

In General Relativity, the formulation by Thorne of the hoop conjecture have as a consequence that cylindrical collapsing matter will not form a black hole. However, the hoop conjecture was given
for spacetimes with a zero cosmological constant. In the presence of a negative cosmological constant one can expect the occurrence of major changes. Indeed, as was shown by Lemos and Zanchin
\cite{lemos} there are black hole solutions with cylindrical symmetry if a negative cosmological constant is present. Charged rotating black string solutions has many similarities with the
Kerr-Newman black hole, apart from space-time being asymptotically anti-de Sitter in the radial direction (and not asymptotically flat). The existence of black strings suggests that they could be
the final state of  the collapse of matter having cylindrical symmetry.

The problems of determining \(\langle\varphi^{2}\rangle_{ren}\) and investigate the renormalized stress tensor components for conformally coupled massless scalar fields in black String backgrounds
were studied by DeBenedictis in \cite{debenedictis1,debenedictis2}, who used the obtained  \(\langle T_{\mu}^{\nu}\rangle_{ren}\) for the calculation of gravitational backreaction of the quantum
field. In our previous papers \cite{owencabo1,owencabo2} we develop the Schwinger-DeWitt approximation for the stress energy tensor for a massive scalar field in the static cylindrical black hole
background, showing that for a range of values of the coupling constant a violation of the weak energy condition can occur at the horizon of the considered space-time.

In this paper we consider the problem of evaluating the large mass limit of the renormalized vacuum expectation values of the Stress-Energy Tensor for a massive spinor field in a background
space-time having cylindrical symmetry. The general results are applied to explicitly evaluate closed expression for those quantities in the special background formed by a neutral and non-rotating
cylindrical Black String. In Section II first we collect some information about spinor field theory on a general curved background. In Section III we develop the Schwinger-DeWitt approximation for
the one-loop effective action of the spinor field taking into account terms up to the second order in the inverse mass of the field and derive analytical expressions for the components of the
renormalized vacuum expectation value of the stress-energy tensor of the quantum field . Section IV is devoted to review the metric tensor which solves the Einstein-Maxwell equations in the
considered cylindric symmetry situation. Finally, employing the explicit form of the Black-String metric, closed expressions for the renormalized components of the Energy-Momentum tensor are derived
in Section IV. These results can be used to study the vacuum polarization and the back-reaction of the quantum spinor field in the gravitational background. The last section contains some conclusive
comments and future directions of this work.

 In the following we use for the Riemann tensor, its contractions, and the covariant derivatives the sign conventions of Misner, Thorne and Wheeler \cite{misner}. Our units are such that
 \(\hbar=c=G=1\).



\section{Free Spinor Field in Curved Space}

Consider a single massive neutral spinor field in a gravitational background with metric tensor \(g_{\mu\nu}\) in four dimensions. The action for the system is::
\begin{equation}\label{}
	S=S_{gravity}+S_{matter}
\end{equation}
with \(S_{gravity}\) the Einstein-Hilbert action for the gravitational background field and \(S_{matter}\) that of the Dirac field:
\begin{equation}\label{}
	S_{gravity}=\frac{1}{16\pi}\int d^{4}x\sqrt{-g}\left(R-2\Lambda\right)
\end{equation}
and:
\begin{equation}\label{}
	S_{matter}=\frac{i}{2}\int d^{4}x\sqrt{-g}\widetilde{\phi}\left[\gamma^{\mu}\nabla_{\mu}\phi+ m\phi\right]\label{class.action}
\end{equation}
where \(m\) is the mass of the field, \(\phi\) provides a spin representation of the vierbein group and \(\widetilde{\phi}=\phi^{*}\gamma\), where * means transpose. The Dirac matrices \(\gamma\)
and\(\gamma^{\mu}\) satisfy the usual relation \(\left[\gamma^{\mu},\gamma^{\nu}\right]_{+}=2g_{\mu\nu}\widehat{I}\), where \(\widehat{I}\) is the \(4\times4\) unit matrix. The covariant derivative
of any spinor \(\zeta\) obey the conmutation relations:
\begin{equation}\label{}
	\nabla_{\mu}\nabla_{\nu}\zeta-\nabla_{\nu}\nabla_{\mu}\zeta=\frac{1}{2}\mathfrak{F}_{[\alpha,\beta]}R^{\alpha\beta}_{\ \ \ \mu\nu}
\end{equation}
\begin{equation}\label{}
	\nabla_{\nu}\nabla_{\sigma}\nabla_{\mu}\zeta-\nabla_{\sigma}\nabla_{\nu}\nabla_{\mu}\zeta=\frac{1}{2}\mathfrak{F}_{[\alpha,\beta]}R^{\alpha\beta}_{\ \ \
	\mu\sigma}\nabla_{\mu}\zeta+\nabla_{\rho}\zeta  R_{\mu \ \ \nu\sigma}^{\ \ \rho}
\end{equation}
\begin{equation}\label{}
   \nabla_{\sigma}\nabla_{\tau}\nabla_{\nu}\nabla_{\mu}\zeta-\nabla_{\tau}\nabla_{\sigma}\nabla_{\nu}\nabla_{\mu}\zeta=\frac{1}{2}\mathfrak{F}_{[\alpha,\beta]}R^{\alpha\beta}_{\ \ \
	\sigma\tau}\nabla_{\nu}\nabla_{\mu}\zeta+\nabla_{\nu}\nabla_{\rho}\zeta  R_{\mu \ \ \sigma\tau}^{\ \ \rho}+\nabla_{\rho}\nabla_{\mu}\zeta  R_{\nu \ \ \sigma\tau}^{\ \ \rho}
\end{equation}
and so forth. In equations (4) to (6) :
\begin{equation}\label{}
	\mathfrak{F}_{\left[\alpha,\beta\right]}=\frac{1}{4}\left[\gamma_{\alpha},\gamma_{\beta}\right]_{-}
\end{equation}
are the generators of the vierbein group, \(\left[\ \ , \ \ \right]_{-}\) is the commutator bracket, and :
\begin{equation}\label{}
	 R^{\alpha\beta}_{\ \ \ \mu\nu}=h^{\alpha}_{\ \sigma}h^{\beta}_{\ \tau}R^{\sigma\tau}_{\ \ \ \mu\nu}
\end{equation}
where \(h^{\alpha}_{\ \beta}\) is the vierbein which satisfies \(h_{\alpha\mu}h^{\alpha}_{\ \nu}=g_{\mu\nu}\). The covariant derivatives of \(\gamma\), \(\gamma^{\mu}\) and
\(\mathfrak{F}_{\left[\alpha,\beta\right]}\) vanishes.

By applying the functional derivative operation to the action (\ref{class.action}) with respect to the spinor field \(\phi\) we obtain the desired equation of motion for the field:
\begin{equation}\label{}
	\left(\gamma^{\mu}\nabla_{\mu}+m\right)\phi=0 \label{diraceqn1}
\end{equation}

\section{Schwinger-DeWitt approximation for the renormalized effective action of the spinor field on a general curved background}

In this section we give exact expressions for the renormalized effective lagrangian, the corresponding renormalized effective action, and the renormalized vaccum expectation value of the stress
energy tensor for the neutral spinor field obeying equation (\ref{diraceqn1}) in the large mass limit. This approximation is known as the Schwinger-DeWitt one, and before applying this approach to
the particular problem considered in this work we make the following remarks.

In the first place, we mention that the Schwinger-DeWitt technique is directly applicable to "minimal" second order differential operators ( acting on the super-field \(\phi^{A}\)) that have the
general form:
\begin{equation}\label{}
	 \hat{D}=\Box-m^{2}+Q \label{minimalop}
\end{equation}
where \(\Box\,=\,g^{\mu\nu}\nabla_{\mu}\nabla_{\nu}\) is the covariant D'Alembert operator, \(\nabla_{\mu}\) is the covariant derivative defined by means of some background conection
\(\mathfrak{C}_{\mu}\left(x\right)\),
\begin{equation}\label{}
	 \nabla_{\mu}\phi^{A}=\partial_{\mu}\phi^{A}+\mathfrak{C}^{A}_{\ B\mu}\phi^{B}\label{covderivative}
\end{equation}
\(g^{\mu\nu}\) is the metric of the background spacetime, \(m\) is the mass parameter of the quantum field and \(Q^{A}(x)\) is an arbitrary matrix playing the role of the potential.

The explicit form of the background affine connection \(\mathfrak{C}^{A}_{\ B\mu}(x)\)that defines the covariant derivative (\ref{covderivative}) is not needed, only is necessary to know the
commutator of covariant derivatives that defines curvature:
\begin{equation}\label{}
	\left[\nabla_{\alpha},\nabla_{\beta}\right]_{-}\phi=\mathfrak{R}_{\alpha\beta}\phi
\end{equation}
\begin{equation}\label{}
	\mathfrak{R}_{\alpha\beta}=\partial_{\alpha}\mathfrak{C}_{\beta}-\partial_{\beta}\mathfrak{C}_{\alpha}+\left[\mathfrak{C}_{\alpha},\mathfrak{C}_{\beta}\right]_{-}
\end{equation}
In the spinor case of our interest the curvature has the form:
\begin{equation}\label{}
	  \mathfrak{R}_{\alpha\beta}=\gamma^{\sigma}\gamma^{\tau}R_{\sigma\tau\alpha\beta} \label{spinorcurvature}
\end{equation}

The usual formalism of Quantum Field Theory give an expression for the effective action of the quantum field \(\phi\) as perturbation expansion in the number of loops:
\begin{equation}\label{}
	\Gamma\left(\Phi\right)=S\left(\Phi\right)+\sum_{k\geq1}\Gamma_{(k)}\left(\Phi\right)
\end{equation}
where \(S\left(\Phi\right)\) is the classical action of the free field. The one loop contribution of the field \(\phi\) to the effective action is expressed in terms of the operator
(\ref{minimalop}) as:
\begin{equation}\label{}
	\Gamma_{(1)}=-\frac{i}{2}\ln\left(s\det\hat{D}\right)
\end{equation}
where \(s\det\hat{F}=\exp(str\ln\hat{F})\) is the functional Berezin superdeterminant \cite{avramidi} of the operator \(\hat{F}\), and:
\begin{equation}\label{}
	str \hat{F}=\left(-1^{i}\right)F^{i}_{i}=\int d^{4}x\left(-1^{A}\right){F}^{A}_{A}(x)
\end{equation}
is the functional supertrace \cite{avramidi}.

Using the Schwinger-DeWitt representation for the Green´s function of the operator (\ref{minimalop}), we can obtain for the renormalized one loop effective action of the quantum field \(\phi\) the
expression:
\begin{equation}\label{}
		 \Gamma_{(1) ren}\,=\,\int  d^{4}x \sqrt{-g}\,\mathfrak{L}_{ren}
\end{equation}
where the renormalized effective Lagrangian reads:
\begin{equation}
\mathfrak{L}_{ren}\,=\,{1\over 2(4\pi)^{2}\,} \sum_{k=3}^{\infty}{\,str \,a_{k}(x,x)\over k(k-1)(k-2)m^{2(k-2)}}\label{renlagrangian},
\end{equation}
The coefficients \([a_{k}]= \,a_{k}(x,x')\), whose coincidence limit appears under the supertrace operation in (\ref{renlagrangian}) are the Hadamard-Minakshisundaram-DeWitt-Seeley coefficients
(HMDS), whose complexity rapidly increases with \(k\). As usual, the first three coefficients of the DeWitt-Schwinger expansion, $a_{0},\,a_{1},\,{\rm and}\,a_{2}, $ contribute to the divergent part
of the action and can be absorbed in the classical gravitational action by renormalization of the bare gravitational and cosmological constants. Various authors have calculated some of the HDSM
coefficients in exact form up to $k \leq 4$ \cite{DeWitt,gilkey,avramidi}.

Now the problem is that the differential operator \(\hat{A}=\gamma^{\mu}\nabla_{\mu}+m\) that appears acting on the spinor field in (\ref{diraceqn1}) is not of the appropiate form (\ref{minimalop}).
this problem is solved if one introduces a new spinor variable \(\psi\) connected with \(\phi\) by the relation \(\phi=\gamma^{\sigma}\nabla_{\sigma}\psi-m\psi\) so that (\ref{diraceqn1}) take the
form:
\begin{equation}\label{}
	 \gamma^{\mu}\gamma^{\nu}\nabla_{\mu}\nabla_{\nu}\psi-^{2}m\psi=0
\end{equation}
Using the properties of Dirac matrices and (\ref{spinorcurvature}) we can establish the identity \(\gamma^{\mu}\gamma^{\nu}\nabla_{\mu}\nabla_{\nu}=\hat{I}\left(\Box-\frac{1}{4}R\right)\) so that
equation (\ref{diraceqn1}) takes the desired form:
\begin{equation}\label{}
	 \left(\Box-\frac{1}{4}R-m^{2}\right)\psi=0 \label{diraceqn2}
\end{equation}
where the potential matrix can be easily identified as \(Q=-\frac{1}{4}R \hat{I}\).

Restricting ourselves here to the terms proportional to $m^{-2},$ using integration by parts and the elementary properties of the Riemann tensor \cite{avramidi,matyjasek}, we obtain for the
renormalized effective lagrangian ,
\begin{equation}\label{}
	\mathfrak{L}_{ren}\,=\,\mathfrak{L}_{ren}^{vac}+\widetilde{\mathfrak{L}_{ren}},
\end{equation}
where \(\widetilde{\mathfrak{L}_{ren}}\) is a portion of the total lagrangian density that vanishes if the background spacetime is a vacuum solution of the classical Einstein equations and
\(\mathfrak{L}_{ren}^{vac}\) is the remaining part of the total renormalized effective lagrangian density of the quantum field. The explicit expressions for this functions in the case of the spinor
field considered in this work reads:
\begin{eqnarray}
 \nonumber
 \widetilde{\mathfrak{L}_{ren}}\,&=&\,{1\over 192 \pi^{2} m^{2}} \left[\,{1\over 28} R_{\mu \nu} \Box R^{\mu \nu}- \frac{3}{280} R
 \Box R\,+\,{1\over 864} R^{3}\,
 \,-\,{1\over 180} R R_{\mu \nu } R^{\mu \nu}\right. \nonumber \\
 &-& {25\over 756} R^{\mu}_{\nu} R^{\nu}_{\gamma} R^{\gamma}_{\mu}
 \,+\,{47\over 1260} R^{\mu \nu}
 R_{\gamma \varrho} R^{\gamma ~ \varrho}_{~ \mu ~ \nu}
\,+\, \,{19\over 1260} R_{\mu \nu} R^{\mu}_{~ \sigma \gamma \varrho} R^{\nu \sigma \gamma \varrho}
\nonumber \\
&-& \left.{7\over 1440}R R_{\mu \nu \gamma \varrho} R^{\mu \nu \gamma \varrho} \right],
\end{eqnarray}
and
\begin{eqnarray}
 \nonumber
 \mathfrak{L}_{ren}^{vac}\,&=&\,{1\over 192 \pi^{2} m^{2}} \left[ {29\over 7560} {R_{\gamma \varrho}}^{\mu \nu} {R_{\mu \nu}}^{\sigma \tau} {R_{ \sigma \tau}}^{\gamma \varrho}\,-\,{1\over 108} R^{\gamma ~ \varrho}_{~ \mu ~ \nu}
 R^{\mu ~ \nu}_{~ \sigma ~ \tau} R^{\sigma ~ \tau}_{~ \gamma ~ \varrho}\right].
\end{eqnarray}

By standard functional differentiation of the effective action with respect to the metric,  the renormalized Stress-Energy tensor is obtained according to the known formula:
\begin{equation}\label{}
	\langle T_{\mu\nu}\rangle_{ren}=\frac{2}{\sqrt{\ -g}}\frac{\delta W_{ren}}{\delta g^{\mu\nu}}
\end{equation}
The result can be written in a general form as
\begin{equation}
   \langle T_{\mu}^{\ \ \nu}\rangle_{ren}=V_{\mu}^{\ \ \nu}+D_{\mu}^{\ \ \nu},
   \label{emTensor}
\end{equation}
where again $D_{\mu}^{\ \ \nu}$ is the part of the total stress tensor that vanishes if the background spacetime is a vacuum solution of the classical Einstein equations and $V_{\mu}^{\ \ \nu}$ is
the remaining part. The above tensors evaluated in this work take the forms
\begin{eqnarray*}
 D_{\mu}^{\ \ \nu} &=&\frac{1}{96\pi^{2} m^{2}}\left[-\frac{3}{280} ( \nabla_{\mu}R\nabla^{\nu}R\,+\,\nabla^{\nu}\nabla_{\mu}(\Box
R)\,+\,\nabla_{\mu}\nabla^{\nu}(\Box R)\,-\,2 \Box^{2} R \delta_{\mu}^{\ \ \nu}\right.
\nonumber \\
&&\left. \,-\ {1\over 2} \delta_{\mu}^{\ \ \nu}\nabla_{\gamma}R\nabla^{\gamma}R \,-\,2 \Box R \nabla^{\nu}\nabla_{\mu}R )\,+\,{1\over 28}\left[\nabla_{\mu}R_{\gamma \lambda}\nabla^{\nu}R^{\gamma
\lambda}\,-\, \nabla^{\nu}R_{\gamma \lambda}\nabla^{\lambda}R^{\ \ \gamma}_{\mu}\right.\right.
\nonumber \\
&&\left.\left.\,-\, \nabla_{\mu}R_{\gamma \lambda}\nabla^{\lambda}R^{\gamma \nu}\,+ \nabla^{\gamma}R_{\gamma \lambda}\nabla^{\nu}R^{\ \ \lambda}_{\mu} \,+\,\nabla^{\gamma}R_{\gamma
\lambda}\nabla_{\mu}R^{\lambda \nu}\,+\, \nabla^{\gamma}\nabla^{\nu}(\Box R_{\gamma \mu})\,-\,\Box^{2} R_{\mu}^{\ \ \nu} \right.\right.
\nonumber \\
&&\left.\left.\,+\nabla^{\gamma}\nabla_{\mu}(\Box R_{\gamma}^{\ \nu}) \,-\, {1\over 2} \nabla_{\varrho}R_{\gamma \lambda}\nabla^{\varrho} R^{\gamma \lambda}\delta_{\mu}^{\ \nu}\,-
\nabla^{\gamma}\nabla^{\lambda}(\Box R_{\gamma \lambda})\delta_{\mu}^{\ \nu}\,+\,\nabla_{\lambda}\nabla^{\nu}R_{\gamma \mu}R^{\gamma \lambda} \right.\right.
\nonumber \\
&&\left.\left.\,+\, \nabla_{\lambda}\nabla_{\mu}R_{\gamma}^{\ \nu} R^{\gamma \lambda}\,-\,
 \left(-\nabla_{\lambda} \nabla_{\sigma} R_{\ \gamma}^{\lambda \ \ \sigma \nu}R^{\ \ \gamma}_{\mu}+{1\over 2} \nabla^{\nu} \nabla_{\gamma}R \,-\, R^{\lambda \sigma}R_{\gamma \ \lambda \sigma}^{\ \nu}\right)R^{\ \ \gamma}_{\mu}
\right.\right.
\nonumber \\
&&\left.\left.\,-\ \nabla^{\gamma}\nabla^{\nu}R_{\gamma \lambda}R^{\ \ \gamma}_{\mu}\,+\, R_{\gamma\lambda}R^{\gamma\lambda}R^{\ \ \gamma}_{\mu} \,-\, \Box R_{\gamma \mu}R^{\gamma \nu} \,-
\nabla^{\lambda}\nabla_{\mu}R_{\gamma \lambda}R^{\gamma \nu}\right]\right.
\nonumber \\
&&\left.\left. \,-\,{25\over 756}\left[ {3\over 2}\nabla^{\nu}R_{\gamma \lambda}\nabla^{\lambda} R^{ \ \ \gamma}_{\mu}\,-\,{3\over 2} \nabla^{\gamma} R_{\gamma \lambda} \nabla^{\varrho}
R_{\varrho}^{ \ \lambda } \delta_{\mu}^{\ \nu}\,-\,{3\over 2} \nabla_{\varrho}R_{\gamma \lambda} \nabla^{\lambda} R^{ \gamma \varrho} \delta_{\mu}^{\ \nu}\,+\,{3\over 2}\nabla_{\lambda}
\nabla^{\nu}R_{\gamma \mu}R^{\gamma \lambda}\right.\right.\right.
\nonumber \\
&&\left.\left.\,+\,{3\over 2} \nabla_{\lambda} \nabla_{\mu}R_{\gamma}^{\ \nu} R^{\gamma \lambda}\,-\, {3\over 2}\nabla^{\lambda} \nabla_{\varrho}R_{\gamma \lambda}R^{\gamma \varrho} \delta_{\mu}^{ \
\nu}\,+\,{3\over 2} \nabla^{\lambda} \nabla^{\nu}R_{\gamma \lambda} R^{ \ \ \gamma}_{\mu} \,-\,{3\over 2}\left(\nabla_{\lambda} \nabla_{\sigma} R_{\ \gamma}^{\lambda \ \ \sigma \nu}R^{\ \
\gamma}_{\mu}\right.\right.\right.
\nonumber \\
&&\left.\left.\left. \,+\,{1\over 2} \nabla^{\nu} \nabla_{\gamma}R \,-\, R^{\lambda \sigma}R_{\gamma \ \lambda \sigma}^{\ \nu}\,+\, R_{\gamma\lambda}R^{\gamma\lambda}\right) R^{\ \
\gamma}_{\mu}+{3\over 2} \nabla^{\gamma} R_{\gamma \lambda}\nabla_{\mu}R^{\lambda \nu}\right.\right.
\nonumber \\
&&\left.\left.\,+\,{3\over 2}\nabla^{\lambda} \nabla_{\mu}R_{\gamma \lambda} R^{\gamma \nu}\,-\, {3\over 2}\Box R_{\gamma \mu} R^{\gamma \nu} \,-\,{3\over 2} \nabla_{\varrho}
\nabla^{\gamma}R_{\gamma \lambda} R^{\lambda \varrho} \delta_{\mu}^{ \ \nu}\,+\,R_{\gamma \lambda} R_{\varrho}^{\ \gamma} R^{\lambda \varrho}\delta_{\mu}^{\ \nu}\right.\right.
\nonumber \\
&&\left.\left. \,-\, 3 R_{\gamma \lambda} R^{\ \ \gamma}_{\mu} R^{\lambda \nu}\,+\,{3\over 2}\nabla_{\mu}R_{\gamma \lambda} \nabla^{\lambda} R^{\gamma \nu} \,-\, 3 \nabla_{\gamma} R_{\gamma \mu}
\nabla^{\lambda} R^{\gamma \nu}\,+\, {3\over 2}\nabla^{\gamma}R_{\gamma \lambda} \nabla^{\nu} R^{\ \ \lambda}_{\mu}\right]\right.
\nonumber \\
&&\left.\,+\,{47\over 1260}\left(\nabla^{\gamma}R_{\gamma \mu}\nabla_{\lambda} R_{\lambda}^{\ \nu}\,+\,\nabla_{\lambda}R_{\gamma}^{\ \nu}\nabla^{\gamma} R^{\ \ \lambda}_{\mu}\,-\, 2
\nabla^{\gamma}R_{\gamma \lambda} \nabla^{\lambda} R_{\mu}^{\ \ \nu}\,-\,\nabla^{\nu} R_{\gamma \lambda} \nabla^{\varrho} R_{\varrho \ \ \mu}^{\ \gamma \lambda} \right.\right.
\nonumber \\
&&\left.\left.\,+\,\nabla_{\varrho}R_{\gamma \lambda} \nabla_{\mu}R^{\gamma \varrho \lambda \nu} \,+\, 2 \nabla_{\varrho}R_{\gamma \lambda} \nabla^{\sigma}R_{\sigma}^{\ \gamma \lambda \varrho }
\delta_{\mu}^{\ \nu}\,-\,\nabla_{\lambda} \nabla_{\gamma} R_{\mu}^{\ \ \nu}R^{\gamma \lambda}\,+\, \nabla^{\varrho} \nabla^{\nu}R_{\gamma \lambda \varrho \mu}R^{\gamma \lambda}\right.\right.
\nonumber \\
&&\left.\left.\,-\,\Box R_{\gamma \mu \lambda}^{\ \ \ \ \nu} R^{\gamma \lambda}\,+\,\nabla^{\varrho} \nabla_{\mu}R_{\gamma \ \lambda \varrho} ^{\ \nu} R^{\gamma \lambda}\,-\,\nabla^{\lambda}
\nabla^{\sigma}R_{\gamma \lambda \varrho \sigma} R^{\gamma \varrho} \delta_{\mu}^{\ \nu}\,+\, {1\over 2} \nabla^{\gamma} \nabla_{\lambda}R_{\gamma}^{\ \ \nu} R^{\ \ \lambda}_{\mu}\right.\right.
\nonumber \\
&&\left.\left.\,+\,{1\over 2} \nabla_{\lambda} \nabla^{\gamma}R_{\gamma}^{\ \nu}R^{ \ \ \lambda}_{\mu} \,+\, {1\over 2} \nabla^{\gamma} \nabla_{\lambda}R_{\gamma \mu} R^{\lambda \nu}\,+\, {1\over 2}
\nabla_{\lambda} \nabla^{\gamma}R_{\gamma \mu} R^{\lambda \nu}\,+\, {1\over 2} R_{\gamma \lambda} R_{\varrho \sigma} R^{\gamma \varrho \lambda \sigma} \delta_{\mu}^{\ \nu} \right.\right.
\nonumber \\
&&\left.\left.\,-\,{3\over 2} R_{\gamma \lambda} R_{\varrho}^{\ \nu} R^{\gamma \varrho \lambda}_{\ \ \ \mu}\,-\, {3\over 2} R_{\gamma \lambda} R_{\varrho \mu} R^{\gamma \varrho \lambda
\nu}\,-\,\nabla^{\gamma} \nabla^{\lambda}R_{\gamma \lambda} R_{\mu}^{\ \ \nu} \,+\, \nabla_{\varrho} \nabla^{\nu}R_{\gamma \lambda} R^{\gamma \varrho \lambda}_{\ \ \ \mu}\right.\right.
\nonumber \\
&&\left.\left.\,+\, \nabla_{\varrho} \nabla_{\mu}R_{\gamma \lambda} R^{\gamma \varrho \lambda \nu}\,-\,\nabla_{\sigma} \nabla_{\varrho }R_{\gamma \lambda} R^{\gamma \sigma \lambda \varrho}
\delta_{\mu}^{\ \nu}\,-\, \Box R_{\gamma \lambda} R^{\gamma \ \ \lambda \nu}_{\ \mu}\,-\, \nabla_{\mu} R_{\gamma \lambda} \nabla^{\varrho} R_{\varrho}^{\ \gamma \lambda \nu}\right.\right.
\nonumber \\
&&\left.\left.\,+\,\nabla_{\varrho}R_{\gamma \lambda} \nabla^{\nu} R^{\ \ \gamma \varrho \lambda }_{\mu}\,-\,2 \nabla_{\varrho} R_{\gamma \lambda} \nabla^{\varrho}R^{\gamma \ \ \lambda \nu}_{\
\mu}\right)+ {19\over 1260}\left(2 \nabla_{\lambda}R_{\gamma \mu} \nabla^{\varrho}R_{\varrho}^{\ \nu \gamma \lambda} \,-\, 2 \nabla^{\gamma}R_{\gamma \lambda} \nabla^{\varrho} R_{\varrho \ \ \mu}^{\
\nu \lambda}\right.\right.
\nonumber \\
&&\left.\left.\,-\,2 \nabla_{\varrho}R_{\gamma \lambda} \nabla^{\lambda} R^{\gamma \ \ \varrho \nu}_{ \ \mu}\,-\,2 \nabla_{\varrho} \nabla^{\gamma} R_{\gamma \ \lambda \mu}^{\ \nu} R^{\lambda
\varrho} \,-\, 2 \nabla^{\gamma} \nabla^{\varrho} R_{\gamma \lambda \varrho}^{\ \ \ \nu} R^{ \ \ \lambda}_{\mu}\,+\,2 \nabla_{\varrho} \nabla_{\lambda} R_{\gamma \mu} R^{\gamma \varrho \lambda
\nu}\right.\right.
\nonumber \\
&&\left.\left.\,+\, R_{\gamma \mu } R_{\lambda \varrho \sigma}^{\ \ \ \nu} R^{\gamma \sigma \lambda \varrho} -2 \nabla_{\lambda}R_{\gamma \mu} \nabla^{\varrho}R_{\varrho}^{\ \gamma \lambda
\nu}\,-\,2 \nabla^{\gamma} \nabla_{\varrho} R_{\gamma \lambda} R^{\lambda \ \ \varrho \nu}_{\ \mu}\right) \right.
\nonumber \\
&&\left. -{1\over 180}\left( \nabla^{\nu}R \nabla^{\gamma} R_{\gamma \mu}\,+\,\nabla_{\mu}R \nabla^{\gamma}R_{\gamma}^{\ \nu} \,+\,2 \nabla^{\nu}R_{\gamma \lambda}\nabla_{\mu} R^{\gamma
\lambda}\,-\, \Box R R_{\mu}^{\ \ \nu}\right.\right.
\nonumber \\
&&\left.\left. \,+\, \nabla_{\gamma}R \nabla^{\nu}R^{\ \ \gamma}_{\mu} \,+\, \nabla_{\gamma}R \nabla_{\mu} R^{\gamma \nu} \,-\, 2 \nabla_{\gamma}R \nabla^{\gamma} R_{\mu}^{\ \ \nu}\,+\ R
\nabla^{\gamma}\nabla^{\nu}R_{\gamma \mu}\,+\,R \nabla^{\gamma}\nabla_{\mu}R_{\gamma}^{\ \nu} \right.\right.
\nonumber \\
\end{eqnarray*}
\begin{eqnarray}
\nonumber \\
&&\left.\left. -R \nabla_{\lambda} \nabla_{\gamma} R_{\ \mu}^{\lambda \ \ \gamma \nu}\,-\,{1\over 2}R \nabla^{\nu} \nabla_{\mu}R \,+\,R R^{\lambda \gamma}R_{\mu \lambda \ \gamma}^{\ \ \ \nu}\,-\,R
R_{\mu\lambda}R^{\lambda\nu}\,-\ 2 \nabla_{\gamma}R \nabla^{\lambda}R_{\lambda}^{\ \gamma} \delta_{\mu}^{\ \nu}\right.\right.
\nonumber \\
&&\left.\left. \,+\nabla^{\nu}\nabla_{\mu}R_{\gamma \lambda} R^{\gamma \lambda}\,+\,\nabla_{\mu}\nabla^{\nu}R_{\gamma \lambda} R^{\gamma \lambda} \,+\, \nabla_{\lambda}\nabla_{\gamma}R R^{\gamma
\lambda} \delta_{\mu}^{\ \nu}\,-\ 2 \Box R_{\gamma \lambda} R^{\gamma \lambda} \delta_{\mu}^{\ \nu} \right.\right.
\nonumber \\
&&\left.\left. \,+\, {1\over 2} R R_{\gamma \lambda} R^{\gamma \lambda} \delta_{\mu}^{\ \nu}\,+\,\nabla^{\nu}\nabla_{\gamma}R R^{\ \ \gamma}_{\mu}\,-\ 2 R R_{\gamma}^{\ \nu} R^{\ \ \gamma}_{\mu}
\,+\,\nabla_{\mu}\nabla_{\gamma}R R^{\gamma \nu}\,-\ R_{\gamma \lambda} R^{\gamma \lambda} R_{\mu}^{\ \ \nu}\right)\right.
\nonumber \\
&&\left.- {7\over 1440}\left(\,+\,4 \nabla_{\gamma} R \nabla^{\lambda} R_{\lambda \mu}^{\ \ \ \gamma \nu}\,+\,4 \nabla_{\gamma}R \nabla^{\lambda} R_{\lambda \ \ \mu}^{\ \nu \gamma}\,+\, 2 R
\nabla^{\gamma}\nabla^{\lambda}R_{\gamma \mu \lambda}^{ \ \ \ \ \nu }  \,+\, 2 R \nabla^{\gamma} \nabla^{\lambda}R_{\gamma \ \lambda \mu}^{\ \nu}\right.\right.
\nonumber \\
&&\left.\left.\,- \ {1\over 2} R R_{\gamma \lambda \varrho \sigma} R^{\gamma \lambda \varrho \sigma} \delta_{\mu}^{\ \nu}\,+\, R_{\mu}^{\ \ \nu} R_{\gamma \lambda \varrho \sigma} R^{\gamma \lambda
\varrho \sigma}\,+\, 2 R R_{\gamma \lambda \varrho}^{\ \ \ \nu} R^{\gamma \lambda \varrho}_{\ \ \ \mu}\,+\, 2 \nabla_{\lambda} \nabla_{\gamma}R R_{\ \mu}^{\gamma \ \ \lambda \nu}\right.\right.
\nonumber \\
&&\left.\left.\,+\,2 \nabla_{\lambda} \nabla_{\gamma}R R^{\gamma \nu \lambda}_{\ \ \ \ \mu}\,-\, 2 \nabla_{\varrho}R_{\gamma \lambda}\nabla^{\varrho} R^{\gamma \lambda} \delta_{\mu}^{\ \nu}\,-\,R
\nabla^{\gamma}\nabla^{\lambda}R_{\gamma \lambda}\delta_{\mu}^{\ \nu}\right)\right.
\nonumber \\
&&\left.-{1\over 144}(\nabla_{\mu}R\nabla^{\nu}R \,+\, R \nabla^{\nu}\nabla_{\mu}R\,+\, {1\over 12} R^{3} \delta_{\mu}^{\ \nu}\,-\,R \Box R \delta_{\mu}^{\ \nu}\,-\,\frac{1}{2} R^{2} R_{\mu}^{\ \
\nu}\,-\,\nabla_{\gamma}R\nabla^{\gamma}R \delta_{\mu}^{\ \nu})\right.
\nonumber \\
\end{eqnarray}
and:
\begin{eqnarray}
V_{\mu}^{\ \ \nu} &=&\frac{1}{96\pi^{2} m^{2}}\left[- {7\over 1440}\left( 2 \Box R_{\gamma \lambda \varrho \sigma} R^{\gamma \lambda \varrho \sigma}\delta_{\mu}^{\ \nu}\,+\,2 \nabla_{\tau}R_{\gamma
\lambda \varrho \sigma}\nabla^{\tau} R^{\gamma \lambda \varrho \sigma } \delta_{\mu}^{\ \nu}\,-\, 2 \nabla^{\nu}R_{\gamma \lambda \varrho \sigma}\nabla_{\mu}R^{\gamma \lambda \varrho \sigma
}\right.\right.
\nonumber \\
&&\left.\left. \,- \nabla^{\nu} \nabla_{\mu}R_{\gamma \lambda \varrho \sigma}R^{ \gamma \lambda \varrho \sigma}\,-\, \nabla_{\mu} \nabla^{\nu}R_{\gamma \lambda \varrho \sigma} R^{ \gamma \lambda
\varrho \sigma}\right) \,+\,{87\over 7560}\left(- 2 \nabla^{\varrho} R_{\gamma \lambda \varrho}^{\ \ \ \nu} \nabla^{\sigma} R_{\sigma \mu}^{\ \ \ \gamma \lambda }\right.\right.
\nonumber \\
&&\left.\left.\,-\,\nabla^{\varrho} \nabla_{\sigma} R_{\gamma \lambda \varrho}^{\ \ \ \nu} R^{\gamma \lambda \sigma}_{ \ \ \ \mu} \,-\, \nabla^{\varrho} \nabla_{\sigma} R_{\gamma \lambda \varrho
\mu} R^{\gamma \lambda \sigma \nu}\,-\, \nabla_{\sigma} \nabla^{\gamma} R_{\gamma \ \lambda \varrho}^{\ \nu} R^{\lambda \varrho \sigma}_{ \ \ \ \mu}\,-\,R_{\gamma \lambda \varrho \mu} R_{\sigma
\tau}^{\ \ \varrho \nu} R^{\gamma \lambda \sigma \tau}\right.\right.
\nonumber \\
&&\left.\left.\,-\,\nabla_{\sigma} \nabla^{\gamma} R_{\gamma \mu \lambda \varrho }R^{\lambda \varrho \sigma \nu}\,+\, {1\over 6} R_{\gamma \lambda \varrho \sigma} R_{\tau \chi}^{\ \ \gamma \lambda}
R^{\varrho \sigma \tau \chi} \delta_{\mu}^{\ \nu}\,-\, 2\nabla_{\sigma} R_{\gamma \lambda \varrho }^{\ \ \ \nu} \nabla^{\varrho}R^{\gamma \lambda \sigma}_{\ \ \ \mu}\right)\right.
\nonumber \\
&&\left.+ {19\over 1260}\left(\,-\,{1\over 2} \Box R_{\gamma \mu \lambda \varrho} R^{\gamma \nu \lambda \varrho}\,+\,\nabla^{\nu}R_{\gamma \lambda \varrho \sigma} \nabla^{\sigma}R^{\gamma \lambda
\varrho}_{\ \ \ \mu}\,-\,\nabla^{\gamma}R_{\gamma \lambda \varrho \sigma} \nabla^{\nu} R^{\lambda \ \ \varrho \sigma}_{\ \mu} \right.\right.
\nonumber \\
&&\left.\left.\,-\, \nabla_{\sigma}R_{\gamma \lambda \varrho \mu} \nabla^{\sigma} R^{\gamma \lambda \varrho \nu}\,-\, {1\over 2} \nabla^{\gamma} R_{\gamma \lambda \varrho \sigma} \nabla^{\tau}
R_{\tau}^{\ \lambda \varrho \sigma} \delta_{\mu}^{\ \nu} \,+\,\nabla_{\sigma} \nabla^{\nu} R_{\gamma \mu \lambda \varrho} R^{\gamma \sigma \lambda \varrho} \right.\right.
\nonumber \\
&&\left.\left.\,-\,{1\over 2} \nabla^{\lambda} \nabla_{\tau}R_{\gamma \lambda \varrho \sigma} R^{\gamma \tau \varrho \sigma} \delta_{\mu}^{\ \nu}\,-\, {1\over 2}\nabla_{\tau} R_{\gamma \lambda
\varrho \sigma} \nabla^{\sigma} R^{\gamma \lambda \varrho \tau} \delta_{\mu}^{\ \nu}\,+\, {1\over 2} \nabla_{\tau} \nabla^{\gamma} R_{\gamma \lambda \varrho \sigma} R^{\lambda \tau \varrho \sigma}
\delta_{\mu}^{\ \nu}\right.\right.
\nonumber \\
&&\left.\left.\,+\, \nabla_{\lambda} \nabla^{\nu} R_{\gamma \lambda \varrho \sigma} R^{\gamma \ \ \varrho \sigma}_{\ \mu}\,-\,{1\over 2} \Box R_{\gamma \ \lambda \varrho }^{\ \nu} R^{\gamma \ \
\lambda \varrho}_{\ \mu}\right)\,-\,{1\over 108}\left(3 \nabla^{\gamma} R_{\gamma \lambda \varrho \mu} \nabla^{\sigma} R_{\sigma}^{\ \varrho \lambda \nu}\right.\right.
\nonumber \\
&&\left.\left.\,+\, {1\over 2} R_{\gamma \lambda \varrho \sigma} R_{\tau \ \chi}^{\ \gamma \ \varrho} R^{\lambda \tau \sigma \chi}\delta_{\mu}^{\ \nu}\,+\,3 \nabla^{\gamma} R_{\gamma \lambda \varrho
\sigma} \nabla^{\sigma} R^{\gamma \nu \varrho}_{ \ \ \ \mu} \,-\,{3\over 2}\nabla_{\sigma} \nabla_{\varrho} R_{\gamma \mu \lambda}^{\ \ \ \ \nu} R^{\gamma \sigma \lambda \varrho}\right.\right.
\nonumber \\
&&\left.\left.\,+\,{3\over 2}\nabla_{\sigma} \nabla^{\varrho} R_{\gamma \ \lambda \varrho }^{\ \nu} R^{\gamma \sigma \lambda }_{\ \ \ \mu}\,+\,{3\over 2}\nabla_{\sigma} \nabla^{\varrho} R_{\gamma
\mu \lambda \varrho} R^{\gamma \sigma \lambda \nu} \,-\, 3 R_{\gamma \lambda \varrho \mu} R_{\sigma \ \tau}^{\ \lambda \ \nu} R^{\gamma \sigma \varrho \tau}\right.\right.
\nonumber \\
&&\left.\left.\,+\,{3\over 2}\nabla_{\sigma} \nabla^{\lambda} R_{\gamma \lambda \varrho }^{\ \ \ \nu} R^{\gamma \ \  \varrho \sigma}_{ \ \mu}\,-\,{3\over 2}\nabla^{\lambda} \nabla^{\sigma} R_{\gamma
\lambda \varrho \sigma} R^{\gamma \ \ \varrho \nu}_{\ \mu}\,+\, {3\over 2}\nabla^{\lambda} \nabla_{\sigma} R_{\gamma \lambda \varrho \mu} R^{\gamma \nu \varrho \sigma}\right.\right.
\nonumber \\
&&\left.\left.\,+\,3 \nabla^{\gamma} R_{\gamma \lambda \varrho \sigma} \nabla^{\sigma} R^{\lambda \ \ \varrho \nu}_{\ \mu}\,-\,{3\over 2}\nabla^{\lambda} \nabla^{\sigma} R_{\gamma \lambda \varrho
\sigma} R^{\gamma \nu \varrho}_{\ \ \ \mu} \,+\,3\nabla_{\sigma} R_{\gamma \lambda \varrho }^{\ \ \ \nu} \nabla^{\lambda}R^{\gamma \ \ \varrho \sigma}_{\ \mu}\right]\right.
\nonumber \\
&&\left.\,-\, {3\over 2}\nabla_{\sigma} \nabla_{\varrho} R_{\gamma \ \lambda \ \mu}^{\ \nu } R^{\gamma \sigma \lambda \varrho}\right)
\nonumber \\
\end{eqnarray}
We should stress that a similar calculation was first performed by Matyjasek for Reissner-Nordstrom space-time in reference \cite{matyjasek} and for a general space-time in \cite{matyjasek1}. Our
results are somewhat different from that of Matyjasek, but in view of the existence of many tensor identities relating the metric tensor, the Riemann tensor, its contractions and its covariant
derivatives, we expect the two results will be equivalent. The formulas obtained in reference \cite{matyjasek1} and those obtained by us, when applied to Schwarshild, Reissner-Nordstrom and Kerr
space-time shows identical results for the components of the renormalized stress tensor. However \cite{matyjasek2}, in the case of the black string space-time, as we will show later in this paper,
there are some minor differences in some numerical coefficients in the results for the massive scalar and spinor fields when compared with that obtained employing the formulas obtained by Matyjasek
in reference \cite{matyjasek}. The origin of this minor differences remain unclear for us at the moment of writing this report.

\section{The cylindrical black hole}
The charged rotating Black String or cylindrical black hole spacetime is an stationary cylindrically symmetric solution of the Einstein-Maxwell equations derived from the action (see Ref.
\cite{lemos}):
\begin{equation}\label{}
	S=S_{gravity}+S_{em},
\end{equation}
where \(S_{gravity}\) is given by (2) and:
\begin{equation}\label{}
	S_{em}=-\frac{1}{16\pi}\int
	d^{4}x\sqrt{-g}F^{\mu\nu}F_{\mu\nu}.
\end{equation}
correspond to the presence of an electromagnetic field described by the Maxwell tensor:
\begin{equation}\label{}
	F_{\mu\nu}=\partial_{\mu}A_{\nu}-\partial_{\nu}A_{\mu},
\end{equation}
\(A_{\mu}\) being the vector potential. The metric element in a cylindrical coordinate system \((x^{0},x^{1},x^{2},x^{3})=(t,\rho,\varphi,z)\) with \(-\infty<t<\infty\), \(0\leq \rho<\infty\),
\(-\infty<z<\infty\), \(0\leq \varphi<2\pi\) adequate for the geometry of interest results.
\begin{multline}\label{}
	\ \ \ \ \ \ \ \ \ \ \ \ \ ds^{2}=-(\alpha^{2}\rho^{2}-\frac{2(M+\Omega)}{\alpha\rho}+\frac{4Q^{2}}{\alpha^{2}\rho^{2}})dt^{2}-\frac{16J}{3\alpha\rho}(1-\frac{2Q^{2}}{(M+\Omega)\alpha\rho})dtd\varphi\\
	+[\rho^{2}+\frac{4(M-\Omega)}{\alpha^{3}\rho}(1-\frac{2Q^{2}}{(M+\Omega)\alpha\rho})]d\varphi^{2} \ \ \ \ \ \ \ \ \ \ \ \ \ \ \ \ \ \ \ \ \ \
	\\+\frac{1}{\alpha^{2}\rho^{2}-\frac{2(3\Omega-M)}{\alpha\rho}+\frac{4Q^{2}(3\Omega-M)}{\alpha^{2}\rho^{2}(\Omega+M)}}d\rho^{2}+\alpha^{2}\rho^{2}dz^{2} \ \ \ \ \ \ \ \ \ \ \ \ \ \ \ \ \ \ \ \ \ \ \ \ \ \ \ \ \ \  \label{ds2}\
\end{multline}
where \(M\), \(Q\), and \(J\) are the mass, charge, and angular momentum per unit length of the string respectively. \(\Omega\) is given by
\begin{equation}\label{}
	\Omega=\sqrt{M^{2}-\frac{8J^{2}\alpha^{2}}{9}}.
\end{equation}
The constant \(\alpha\) is defined as follows:
\begin{equation}\label{}
	\alpha^{2}=-\frac{1}{3}\Lambda,
\end{equation}
where \(\Lambda\) is a negative  Cosmological Constant. The corresponding metric element for the static spacetime follow form to relation (\ref{ds2}):
\begin{equation}\label{}
	ds^{2}=-(\alpha^{2}\rho^{2}-\frac{4M}{\alpha\rho})dt^{2}+
	\frac{1}{(\alpha^{2}\rho^{2}-\frac{4M}{\alpha\rho})}d\rho^{2}+
	\rho^{2}d\varphi^{2}+\alpha^{2}\rho^{2}dz^{2}.  \label{ds2simp}
\end{equation}
As we can see from (\ref{ds2simp}),  the considered metric has an event horizon located at \(\rho_{H}=\frac{\sqrt[3]{4M}}{\alpha}\). The apparent singular behavior at this horizon is a coordinate
effect and not a true one. The only true singularity is a polynomial one at the origin, as can it be seen after  calculating the Kretschmann scalar, that results in
\begin{equation}\label{}
	K=R_{\alpha\beta\xi\gamma}R^{\alpha\beta\xi\gamma}=
	24\alpha^{4}\left(1+\frac{M^{2}}{\alpha^{6}\rho^{6}}\right).
\end{equation}

\section{Renormalized Stress-Energy tensor for spinor fields in a
Black String background}

In the space-time of a static Black String metric given by (\ref{ds2simp}) simple results were obtained for the renormalized Stress Tensor of massive spinor field coupled to the background
gravitational field. After a direct calculation, we evaluate for the total stress tensor:
\begin{equation}\label{}
	T_{t}^{\ t}=\frac{1}{40320 \pi^{2}m^{2}\alpha^{3}\rho^{9}}\left(719\alpha^{9}\rho^{9}
	-2976\alpha ^{3}M^{2}\rho^{3}+19072M^{3}\right),
\end{equation}
\begin{equation}\label{}
	T_{z}^{\ z}=T_{\varphi}^{\ \varphi}=\frac{1}{40320 \pi^{2}m^{2}\alpha^{3}\rho^{9}}
	\left(719\alpha^{9}\rho^{9}-3552\alpha
	^{3}M^{2}\rho^{3}+28288M^{3}\right),
\end{equation}
\begin{equation}\label{}
	T_{\rho}^{\ \rho}=\frac{1}{40320 \pi^{2}m^{2}\alpha^{3}\rho^{9}}
	\left(719\alpha^{9}\rho^{9}+3360\alpha
	^{3}M^{2}\rho^{3}-6272M^{3}\right).
\end{equation}
It is interesting to evaluate the above components of the stress tensor at the event horizon of the black string. We obtain the following very simple results:
\begin{equation}\label{}
	T_{t}^{\ t}|_{horizon}=T_{\rho}^{\ \rho}|_{horizon}=2.06\cdot10^{-2}\frac{\alpha^{6}}{\pi^{2}m^{2}}
	\label{horizon1}
\end{equation}
\begin{equation}\label{}
	T_{z}^{\ z}|_{horizon}=T_{\varphi}^{\ \varphi}|_{horizon}=42.9\frac{\alpha^{6}}{\pi^{2}m^{2}}\label{horizon2}
\end{equation}

At this point is useful to make an analysis of the violations or not of any of the well defined Energy Conditions that it is expected to be satisfied for any classical form of matter. We recall in
the following the statements of the main Energy Conditions as appear in the literature \cite{visser}:

\emph{\textbf{Weak energy Condition}}: The weak energy condition states that the energy density of any matter distribution, as measured by any observer in spacetime, must be nonnegative. Because an
observer with four velocity $V^{\alpha}$ measures the energy density to be $T_{\alpha\beta}V^{\alpha}V^{\beta}$, we must have:
\begin{gather}
   T_{\alpha\beta}V^{\alpha}V^{\beta}\geq0
\end{gather}
for any future-directed timelike vector $V^{\alpha}$. This condition implies (in terms of energy density $\varepsilon$ and principal pressures $p_{i}$) that:
\begin{gather}
   \varepsilon\geq0  \ \ \ \ \ \ \ and \ \ \ \ \ \ \ \forall i, \ \ \varepsilon+p_{i}\geq0
\end{gather}
\emph{\textbf{Null energy Condition}}: The null energy condition makes the same statement as the weak form, except that $V^{\alpha}$ is replaced by an arbitrary, future-directed null vector
$k^{\alpha}$. Thus:
\begin{gather}
   T_{\alpha\beta}k^{\alpha}k^{\beta}\geq0
\end{gather}
is the statement of the null energy condition. This condition implies:
\begin{gather}
  \forall i, \ \ \varepsilon+p_{i}\geq0
\end{gather}
\emph{\textbf{Strong energy Condition}}: The statement of the Strong Energy Condition is:
\begin{gather}
   \left(T_{\alpha\beta}-\frac{1}{2}Tg_{\alpha\beta}\right)V^{\alpha}V^{\beta}\geq0
\end{gather}
where $V^{\alpha}$ is any future-directed, normalized, timelike vector $V^{\alpha}$. The strong energy condition therefore implies:
\begin{gather}
   \forall i, \ \ \varepsilon+p_{i}\geq0\ \ \ \ \ \ \ and \ \ \ \ \ \ \ \varepsilon+\sum_{i}p_{i}\geq0
\end{gather}
By virtue of the Einstein field equations,the strong energy condition is really a statement about the Ricci tensor.
\emph{\textbf{Dominant energy Condition}}: The dominant energy condition embodies the notion that matter should flow along timelike or null world lines. Its precise statement is that if $V^{\alpha}$
is an arbitrary, future-directed, timelike vector field, then:
\begin{gather}
   T_{\alpha\beta}V^{\alpha}V^{\beta}\geq0\ \ \ \ \ \ \ and \ \ \ \ \ \ \ -T^{\alpha}_{\beta}V^{\beta}\ \ \ \ \ \ \ is\ not\ spacelike
\end{gather}
for any future-directed timelike vector $V^{\alpha}$. This condition implies (in terms of energy density $\varepsilon$ and principal pressures $p_{i}$) that:
\begin{gather}
   \varepsilon\geq0  \ \ \ \ \ \ \ and \ \ \ \ \ \ \ \forall i, \ \ \varepsilon\geq\mid p_{i}\mid
\end{gather}
All the energy conditions mentioned above are local conditions. There exist some other averaged energy conditions that are not of importance to this work. Both local and averaged energy conditions
play an essential role in the formulation of classical singularity theorems as the Penrose and the Hawking-Penrose singularity theorems \cite{hawking-ellis}, that invokes the weak and strong energy
conditions respectively for their proof. Also the proof of the zeroth law of black hole thermodynamics (the constancy of the surface gravity over the event horizon) relies on the dominant energy
condition and the proof of the second law of black hole thermodynamics (the area increase theorem) uses the null energy condition \cite{wald}. While the classical validity of the energy conditions
are perfectly reasonable assumptions, semiclassical quantum effects are capable of violating the null, weak, strong, and dominant energy conditions.

As can be easily seen from (\ref{horizon1}), the quantum spinor field violates the weak energy condition at the horizon of the black string: the energy density (defined as \(\varepsilon=-T_{t}^{\
t}\)) is negative on this hypersurface. Further analysis also reveals that at the horizon the dominant energy condition is violated as well. At this point it is interesting to compare the results
obtained for the energy conditions in the massive scalar \cite{owencabo1,owencabo2} and the massless scalar \cite{debenedictis2}field cases. The results for the horizon values of the large mass
limit of the renormalized stress tensor for massive scalar field on the black static string background reads:
\begin{equation}\label{}
	T_{t}^{\ t}|_{horizon}=-\frac{3}{2}\frac{\alpha^{6}\eta}{\pi^{2}m^{2}}\left(\frac{1}{40}+3\eta^{2}\right)+\frac{\alpha^{6}}{140\pi^{2}m^{2}}
\end{equation}
\begin{equation}\label{}
	T_{z}^{\ z}|_{horizon}=T_{\varphi}^{\ \varphi}|_{horizon}=-\frac{3}{2}\frac{\alpha^{6}\eta}{\pi^{2}m^{2}}\left(\frac{1}{20}+3\eta^{2}\right)+\frac{\alpha^{6}}{112\pi^{2}m^{2}}
\end{equation}
\begin{equation}\label{}
	T_{\rho}^{\ \rho}|_{horizon}=-\frac{3}{2}\frac{\alpha^{6}\eta}{\pi^{2}m^{2}}\left(\frac{1}{40}+3\eta^{2}\right)+\frac{\alpha^{6}}{140\pi^{2}m^{2}}
\end{equation}
In the general case, all the components of the renormalized stress energy tensor of the quantized massive scalar field will be positive at the horizon for the values of the coupling constant
satisfying the relation \(3\eta^{3}+\frac{1}{40}\eta < \frac{1}{210}\). There are some particular cases in which the above relation is always satisfied. The simplest case of the conformal coupling
\(\xi=\frac{1}{6}\) and the minimal one are two important examples. Also for the case \(\xi<\frac{1}{6}\) the components of the quantized scalar field at the horizon are always positive quantities.
Then we can conclude that for the particular cases mentioned above the weak energy condition is violated. In the massless scalar field case it is concluded in \cite{debenedictis2} that the weak,
null and strong energy conditions are violated. The precise conditions under which quantum effects in four dimensional spacetime are capable of violating the averaged energy conditions is unknown.
The quantum-induced violations of the null, weak, strong, and dominant energy conditions are typically very small. After all, by definition these are order $\hbar$ effects. It is far from clear
whether or not it is possible to get a large violation of the energy conditions.


\section{Concluding remarks}
The quantization of a massive  spinor field  coupled to an arbitrary gravitational background space-time  was considered. The renormalized quantum mean values of the Lagrangian and the corresponding
components of the Energy-Momentum tensor are obtained in the large mass limit using the Schwinger-DeWitt approximation up to the second order in the inverse mass of the field, and are explicitly
evaluated in the case of the static black string space-time. The quantum spinor field violates the weak as well as dominant energy conditions at the horizon hypersurface of the black string. This
results are expected to be employed in future works to investigate the back-reaction of the quantum spinor field, on the Black String metric.
\begin{acknowledgements}
One of the authors (O.P.F.P) greatly acknowledges M. Chac\'{o}n Toledo for helpful discussions. We are very much grateful by the helpful remarks of Dr. A.
 Follaci and Dr. J. Matyjasek, and in addition  deeply acknowledge our colleague the MSc. A. Ulacia for a useful introduction
 to computer algebraic evaluations in General relativity.  We also
 ought to express our gratitude by the kind support granted by the Office of
 External Activities (OEA) of the AS ICTP,  under the ICTP Network
 Net-35.
\end{acknowledgements}

\end{document}